\documentclass[aps,prd,twocolumn,twoside,superscriptaddress,floatfix, nofootinbib]{revtex4-1}
\pdfoutput=1
\usepackage{graphicx}
\usepackage{color}
\usepackage{hyperref}
\usepackage{float}
\usepackage{aas_macros}
\usepackage{ifthen,amsmath,amssymb, bm}

\newcommand{\beq} {\begin{equation}}
\newcommand{\eeq} {\end{equation}}
\newcommand{\bal} {\begin{aligned}}
\newcommand{\eal} {\end{aligned}}

\newcommand{\bl}{\boldsymbol{l}}
\newcommand{\bL}{\boldsymbol{L}}

\newcommand{\bll}{\boldsymbol{L}}
\newcommand{\barr}{\begin{eqnarray}}
\newcommand{\earr}{\end{eqnarray}}

\newcommand{\dotfac}[1]{({\bll} \cdot {\bl}_{#1})}
\newcommand{\vsp}{\vphantom{\Big[}\\}

\newcommand{\intL}{\int_{\substack{\bl_1 + \bl_2 \\ =\bll }}}
\newcommand{\intLp}{\int_{\substack{\bl_3 + \bl_4 \\ =\bll' }}}

\begin{document}

\author{Abhishek S. Maniyar}
\email{abhishek.maniyar@nyu.edu}
\affiliation{Center for Cosmology and Particle Physics, Department of Physics, New York University, 726 Broadway, New York, NY, 10003, U.S.A.}
\author{Emmanuel Schaan}
\email{eschaan@lbl.gov}
\affiliation{Lawrence Berkeley National Laboratory, One Cyclotron Road, Berkeley, CA 94720, USA}
\affiliation{Berkeley Center for Cosmological Physics, Department of Physics,
University of California, Berkeley, CA 94720, USA}
\author{Anthony R. Pullen}
\email{anthony.pullen@nyu.edu}
\affiliation{Center for Cosmology and Particle Physics, Department of Physics, New York University, 726 Broadway, New York, NY, 10003, U.S.A.}
\affiliation{Center for Computational Astrophysics, Flatiron Institute, New York, NY 10010, U.S.A.}

\title{
A new probe of the high-redshift Universe:
nulling CMB lensing with interloper-free ``LIM-pair'' lensing
}

\begin{abstract}

Intermediate redshifts between galaxy surveys and the cosmic microwave background (CMB) remain unexplored territory.
Line intensity mapping (LIM) offers a way to probe the $z\gtrsim 1$ Universe, including the epoch of reionization and the dark ages.
Via exact nulling of the lensing kernel,
we show that LIM lensing, in combination with galaxy (resp., CMB) lensing,
can uniquely probe the $z\gtrsim 1$ (resp., pre-reionization) Universe.

However, LIM foregrounds are a key hurdle to this futuristic technique.
While continuum foregrounds can be controlled by discarding modes perpendicular to the line of sight (low $k_\parallel$ modes), interloper foregrounds haven't been addressed in the context of LIM lensing.
In this paper, we quantify the interloper bias to LIM lensing for the first time, and derive a ``LIM-pair'' estimator which avoids it exactly after cross-correlating with CMB lensing.
This new quadratic lensing estimator works by combining two intensity maps in different lines, from the same redshift, whose interlopers are uncorrelated.
As a result, this foreground avoidance method is robust to even large changes in the amplitude of the interloper power and non-Gaussianity.
The cross-spectrum of the LIM-pair estimator with CMB lensing is thus robust to the currently large theoretical uncertainties in LIM modeling at high redshift.

\end{abstract}

\maketitle

\section{Introduction}

The properties of the observable Universe are precisely constrained at redshift $z=1100$ by observations of the cosmic microwave background (CMB) \cite{Planck18I}, and at $z\lesssim 1$ by galaxy surveys.
In between, line intensity mapping (LIM) is a promising approach to fill the gap and study galaxy evolution and cosmology \cite{Kovetz17}. 
Several promising lines like HI (21 cm), Ly-$\alpha$ (121.6 nm), H$\alpha$ (656.28 nm), [CII] (158 $\mu$m), CO 1-0 (2.6 mm) etc. are being targeted by the ongoing and upcoming LIM experiments to map out the  3D large-scale structure (LSS) of the Universe at high redshift.
However, some periods of the Universe's history, such as the Dark Ages when it was mostly neutral, will remain very challenging to probe.
For instance, probing the Dark Ages with 21cm will require peering through overwhelmingly large foregrounds \cite{Haslam82, Rengelink97, Santos05}.

The lensing of the CMB contains information about the high-redshift Universe, including the epoch of reionization and the dark ages \cite{Lewis06}, and will be measured to sub-percent precision by upcoming experiments \cite{SO19, CMBS419}.
However, the contribution to CMB lensing from eg., the Dark Ages, is dwarfed by that from the low-redshift ($z\lesssim 1$) Universe.
Subtracting this low-redshift contribution could in principle be done with tracers of the matter density (galaxy surveys and LIM surveys) \cite{McCarthy21}, however these would need to overlap on the sky and span the whole redshift range between $z=0$ to the redshift of reionization, without any gap.
This therefore appears unfeasible in practice.

Instead, a futuristic approach could be to reconstruct lensing from a LIM survey \cite{Zahn06, Pourtsidou14, Pourtsidou15, Pourtsidou16, Schaan18, Foreman18, Chakraborty19, Feng19} at high redshift, e.g., $z= 5$.
Combining LIM lensing with galaxy shear at $z= 1$, such as from the Rubin Observatory\footnote{\url{http://www.lsst.org}} \cite{LSSTScienceBook}, one can exactly null the contribution of $z \leq 1$ to the LIM lensing, thus delivering a unique probe of the matter distribution at $z=1-5$.
This redshift range is extremely difficult to probe any other way.
Combining instead LIM lensing with CMB lensing at $z=1100$, one can selectively extract the projected matter density field at $z=5-1100$, covering the epoch of reionization, cosmic dawn and the dark ages.
Again, this redshift range is difficult to observe any other way, and doing so with lensing would enable testing how much of the fluctuations in future 21 cm maps during reionization/the dark ages arise from density fluctuations as opposed to ionization or spin-temperature variations (see \cite{Doux16} for an analogous approach with CMB lensing and the Lyman-$\alpha$ forest).
To do this, we extend the so-called ``nulling'' method from the galaxy lensing tomography literature \cite{Huterer05, Bernardeau14, Barthelemy20}, and we generalize it to LIM lensing and CMB lensing below.
This method allows to not only suppress, but instead exactly null, the otherwise dominant low-redshift contribution to the lensing kernels.

LIM lensing has other applications, beyond enabling lensing tomography at high redshift.
For instance, continuum foregrounds typically render the modes perpendicular to the line of sight (LOS), i.e. with $k_\parallel \simeq 0$, unusable for cosmology.
This can prevent us from measuring the cross-correlation of LIMs with 2D fields, such as CMB lensing.
However, by reconstructing the lensing from LIMs, one obtains a field, $\hat{\kappa}_\text{LIM}$, where the modes with $k_\parallel \simeq 0$ are present, enabling cross-correlations with 2D fields like CMB lensing \cite{Foreman18, Schaan18}.
This therefore offers an alternative to tidal reconstruction \cite{Foreman18, Zhu18}, in order to enable these cross-correlations.

The prospect of measuring LIM lensing remains futuristic, because of several challenges.
Recent work \citep[e.g.][]{Foreman18, Schaan18} has shown that the non-Gaussian nature of LIMs (due to non-linear gravitational evolution at low redshifts) biases LIM lensing.
This bias can be avoided or subtracted to some extent with ``bias hardening'' \cite{Foreman18}, a method inspired from CMB lensing \cite{Osborne14, Namikawa13, Planck13XVII, Sailer20} which makes use of our knowledge of the LIM non-Gaussianity.

Another major challenge to LIM lensing is the fact that the observed LIMs are contaminated by foregrounds.
Continuum foregrounds like the cosmic infrared background (CIB) or Milky-Way emission can be highly dominant over the target line signal.
Thanks to their smooth spectral energy distributions, continuum foregrounds can typically be avoided by discarding the 3D Fourier modes with low $k_\parallel$, i.e. almost perpendicular to the line of sight (LOS).
However, line interlopers cannot be avoided in this way. These are galaxies at a different redshift, emitting in a different line which redshifts to the same observed frequency as the target line.
Methods exist to remove part of the interloper contamination, or to quantify it (see \cite{Kovetz17, Pullen13} for a summary). Methods like
bright voxel masking \cite{Gong14, Breysee15, Yue15, Silva15, Sun18}, secondary line identification \cite{Cheng20}, spectral deconfusion \cite{Cheng20},  cross-correlating the LIM with a template of the contaminant \cite{Silva15} alleviate the issue.
Measuring the anisotropy in the 3D power spectrum, analogous to the Alcock-Paczynski effect \cite{Visbal10, Cheng16, Liu16, Lidz16, Gong20}, allows to quantify the residual contamination.
While these methods reduce the amount of interloper emission, they do not completely remove them.

In this paper, we quantify the bias to LIM lensing from interlopers for the first time, and propose a new method to avoid them entirely, without any assumption other than their redshifts.
We derive a new ``LIM-pair'' quadratic estimator for LIM lensing, relying on a pair of LIMs, from two lines $X$ and $Y$ emitted at the same redshift but with uncorrelated interloper foregrounds.
This method is analogous in spirit to the gradient-cleaned estimators of CMB lensing \cite{Madhavacheril18, Darwish21}.
We forecast the signal-to-noise ratio for this estimator for one example line pair.
We compute the various foreground biases to its auto-spectrum, and show that its cross-spectrum with CMB lensing is exactly free of LIM foregrounds.
Furthermore, this cross-correlation of LIM line-pair lensing with CMB lensing has higher SNR than the auto-power spectrum of the LIM-line pair lensing, making it the first one to be detectable in the future.
This paper constitutes a step towards bias-free lensing reconstruction from LIM. 
The ``nulling'' method, applied to LIM-pair lensing, constitutes a new potential probe of the Dark Ages in the future.

\section{Lensing tomography and ``Nulling''}

Similarly to galaxies and the CMB, LIMs constitute source images, emitted at a cosmological distances from us, which are lensed by all the intervening matter distribution in-between.
In the weak lensing regime and the Born approximation, this lensing is entirely determined by one scalar field for each source image (LIM or CMB), the lensing convergence $\kappa$.
In all cases, the lensing convergence is a projection of the matter overdensity field along the line of sight (LOS),
\beq
\kappa(\vec{n}) = \int d\chi \, W_\kappa(\chi) \, \delta (\chi \vec{n}, z(\chi)),
\label{eq:kappa_from_delta}
\eeq
weighted by the lensing kernel $W_\kappa$.
For an image source at a single redshift or distance $\chi_S$, the lensing kernel is given by
\beq
W_{\kappa} (\chi, \chi_S) = \frac{3}{2} \left( \frac{H_0}{c} \right)^2 \frac{\Omega_m^0}{a} \; \chi \left( 1 - \frac{\chi}{\chi_S} \right).
\label{eq:lensing_kernel_single_source}
\eeq
Here $H_0$ and $\Omega_m^0$ are the Hubble parameter and the matter fraction today, $c$ is the speed of light, $a$ is the scale factor, $\chi_S$ the distance of the source (image being lensed) and $\chi$ the distance of the lens (mass causing the lensing).
This lensing kernel is appropriate for CMB lensing, where the source redshift is $z=1100$, and for a thin redshift slice of LIM.
For extended source redshift distributions $dn/dz_S$, e.g., for a galaxy lensing tomographic bin or a LIM with a large redshift coverage, the lensing kernel is simply the redshift-average of the single-source lensing kernel, weighted by the source redshift distribution: 
\beq
W_{\kappa} (\chi) = \int dz_S \; \frac{1}{n}\frac{dn}{dz_S} \; W_\kappa (\chi, \chi(z_S)),
\eeq
where 
$n \equiv \int dz_S\ dn/dz_S$.
In this paper, we consider LIMs coming from a single redshift, or a thin redshift slice, making this last integral unnecessary.
In practice though, LIM lensing analyses will likely be performed in 3D \cite{Foreman18, Chakraborty19}, in order to discard the low $k_\parallel$ modes most affected by continuum foregrounds.
In what follows, we will therefore not address the question of contamination from continuum foregrounds, and we will assume that this problem is solved by the $k_\parallel$ cuts applied to the LIM.
In what follows, we derive the new ``LIM-pair'' estimator in 2D rather than 3D, to avoid technical distractions.
We also do not implement the bias-hardening weights.
Our 2D estimator generalizes trivially to 3D and to the bias-hardening case, while keeping insensitivity to interloper foregrounds, since it relies on using a pair of lines.

From Eq.~\eqref{eq:kappa_from_delta}, we infer all the auto- and cross-spectra of LIM lensing, galaxy lensing and CMB lensing, in the flat sky and Limber approximations:
\beq
C_\ell^{\kappa \kappa'}
=
\int d\chi \;
\frac{W_\kappa(\chi) W_{\kappa'}(\chi)}{\chi^2}
P_m\left( k=\frac{\ell+1/2}{\chi}, z(\chi)\right).
\eeq

\begin{figure}[h]
\centering
\includegraphics[width=\columnwidth]{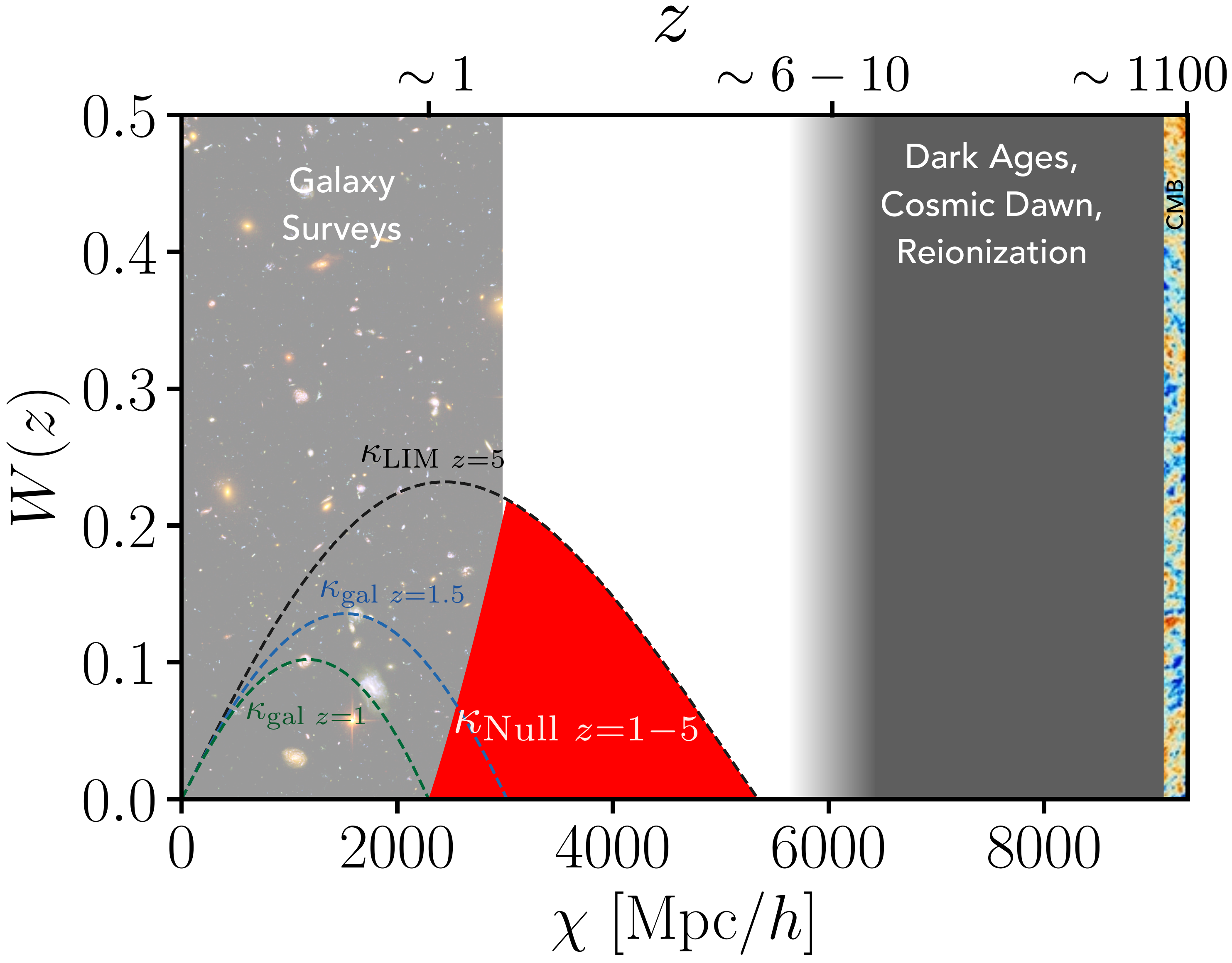}
\includegraphics[width=\columnwidth]{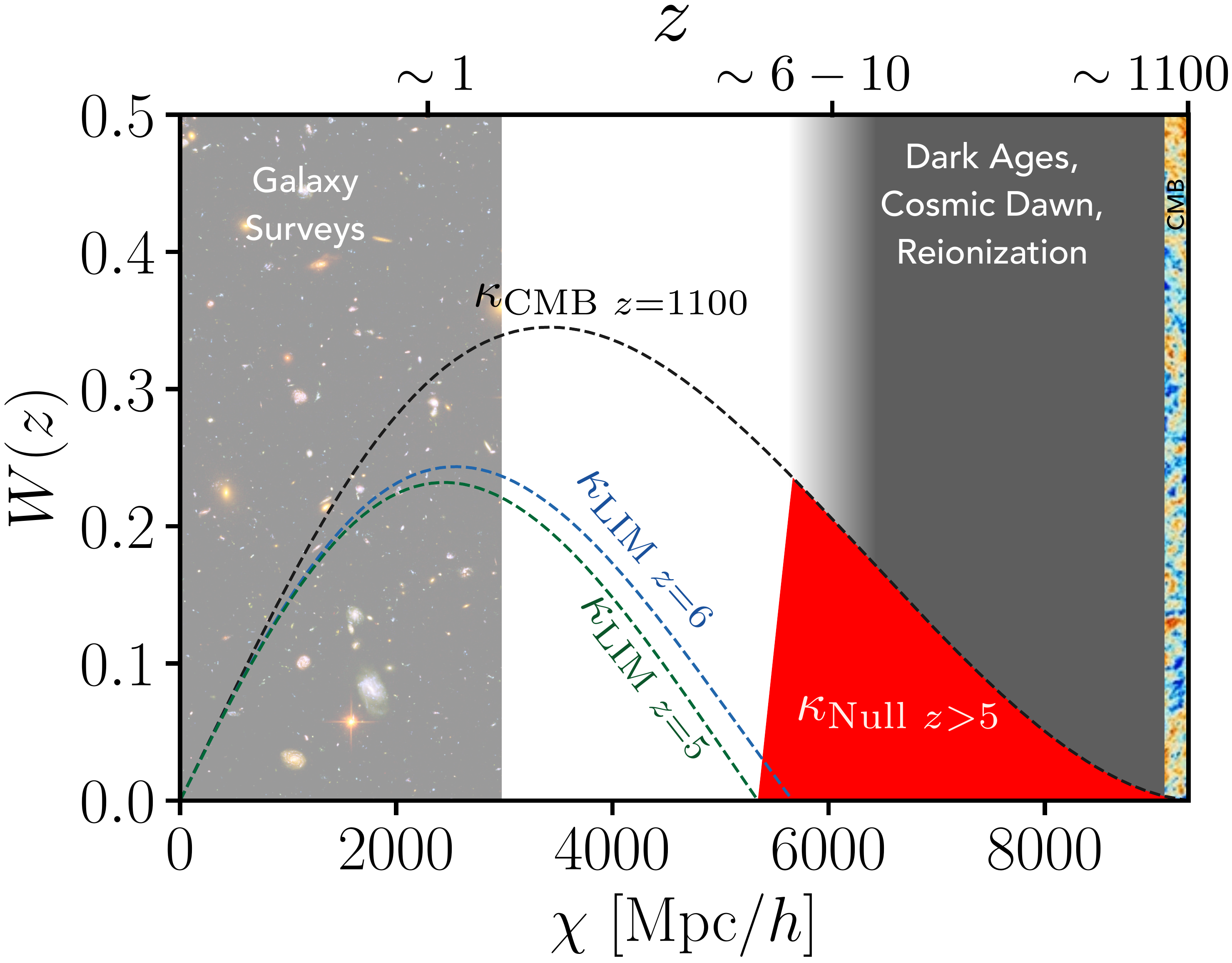}
\centering 
\caption{
While the Universe's properties are very well constrained at low redshift from galaxy surveys and at high redshift with the CMB, many parts of its history remain unexplored.
Top: By combining LIM lensing (dashed black) at $z=5$ with galaxy lensing at $z=1, 1.5$ (dashed blue and green), we construct a linear combination sensitive only to $z=1-5$.
Bottom: By combining CMB lensing (dashed black) and lensing from two LIMs (e.g.,from $z=5$ in green and $z=6$ in blue), one can construct a linear combination which exactly nulls the signal from low redshift ($\kappa_\text{Null}$ in red).
This offers a potential new probe of the Dark Ages, complementary to 21~cm.
However, achieving these futuristic goals requires controlling the foregrounds in LIM, which is the goal of this paper.
}
\label{fig:schematic_nulling}
\end{figure}

As shown in Fig.~\ref{fig:schematic_nulling} for CMB lensing and LIM lensing at redshifts 5 and 6, these lensing kernels span the whole redshift range between the source and the observer.
However, interestingly, Eq.~\eqref{eq:lensing_kernel_single_source} shows that the lensing kernels have a very simple dependence on the lens distance $\chi$: apart from the common overall scale factor, they are second order polynomials in $\chi$.
Such a polynomial is only determined by three coefficients.
An appropriate linear combination of three lensing kernels is therefore sufficient to null these three coefficients, thereby exactly nulling the combined lensing kernel out to the redshift of the closest source \cite{Huterer05, Bernardeau14, Barthelemy20}.
More specifically, for three sources at distances 
$\chi_1 < \chi_2 < \chi_3$,
the linear combination 
\beq
W_{\kappa}(\chi, \chi_3)
+ \alpha  W_{\kappa}(\chi, \chi_2)
- (1+\alpha)  W_{\kappa}(\chi, \chi_1)
\eeq
with
\beq
\alpha=\frac{1/\chi_3 - 1/\chi_1}{1/\chi_1 - 1/\chi_2}
\eeq
is mathematically null for 
$\chi \leq \chi_1$.
In other words, the linear combination
$\kappa_3
+ \alpha \kappa_2
- (1+\alpha)  \kappa_1$
is only sensitive to the matter distribution from $\chi > \chi_1$.

Fig.~\ref{fig:schematic_nulling} illustrates two applications of the nulling method, using LIMs at high redshift.
First, we use one LIM at $z=5$ and two galaxy lensing tomographic bins at $z=1, 1.5$ from e.g., Rubin Observatory.
The nulling combination of these three allows to exactly null any contribution to lensing from $z\leq 1$, providing a probe of the $z=1-5$ Universe.
This probe is valuable because of its redshift range, difficult to access otherwise.
Because this gives the projected matter density field directly, it avoids the need to model the galaxy-halo connection (e.g., galaxy bias).

The second application shown in Fig.~\ref{fig:schematic_nulling} uses two LIMs at $z=5,6$ and CMB lensing.
The nulling combination allows to extract selectively the $z=5-1100$ Universe, exactly nulling any contribution from $z\leq 5$.
This disentangles the contribution from the dark ages, cosmic dawn and the epoch of reionization from the otherwise-dominant low-redshift Universe, yielding a unique probe of the pre-reionization Universe.

In either case, whether we construct $\kappa_\text{Null}$ from LIM and galaxy lensing, or from LIM and CMB lensing, we will be cross-correlating $\kappa_\text{Null}$ with CMB lensing. 
Indeed, the CMB lensing kernel fully overlaps with the nulled lensing kernel, such that $\langle \kappa_\text{Null} \kappa_\text{CMB} \rangle$ is non-zero and probes the same exact redshift range as $\kappa_\text{Null}$.
Furthermore, we will show that this combination is free of interloper bias, when LIM lensing is measured with the LIM-pair estimator.

In the rest of this paper, we focus on a necessary step towards this futuristic prospect: suppressing interloper contamination in LIM.
We show that cross-power spectrum of the form 
$C_L^{\hat{\kappa}_\text{LIM} \hat{\kappa}_\text{CMB}}$
can be measured without interloper bias, thanks to the LIM-pair estimator.
As a result, the cross-spectrum
$C_L^{\hat{\kappa}_\text{Null} \hat{\kappa}_\text{CMB}}$
can also be measured free of interloper bias.
These cross-spectra probe exclusively the high-redshift Universe.
In what follows, we focus on CMB lensing rather than galaxy lensing, but all the results apply identically.

\section{Interloper emission and line pairs}

Throughout this paper, we consider two different lines with widely separated rest-frame frequencies.
We denote by $X$ and $Y$ intensity maps in these two target lines, from galaxies at the same redshift.
Since $X$ and $Y$ trace the large-scale structure distribution of matter at the same redshift, they are correlated and have a non-zero cross-spectrum $C_l^{XY}$.
The two intensity maps $X$ and $Y$ are affected by interloper foregrounds.
However, we assume that the target lines and redshift of $X$ and $Y$ have been selected such that their interlopers do not originate from the same redshift, and are therefore statistically independent.

\begin{figure}[h!]
\centering
\includegraphics[width=\columnwidth]{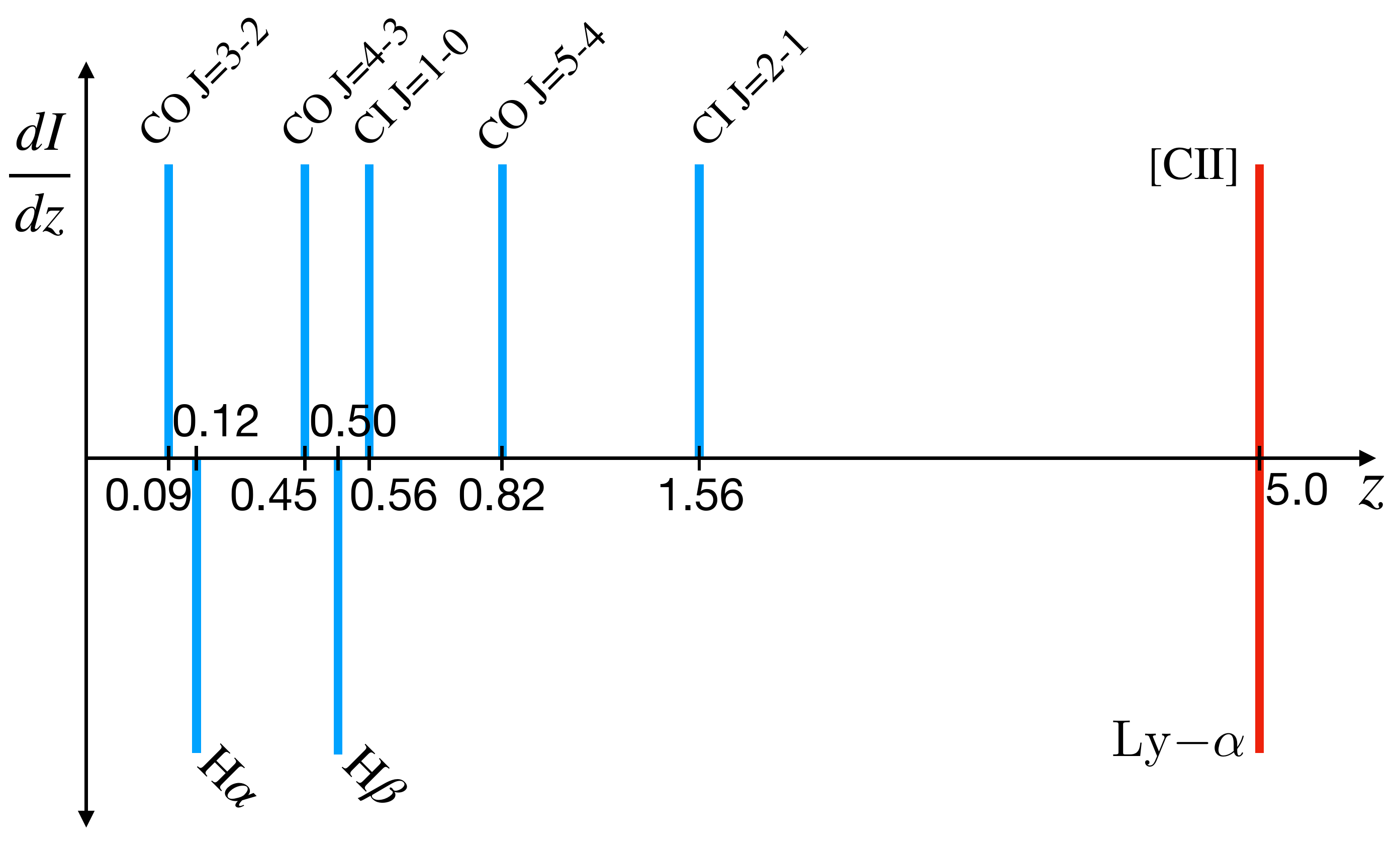}
\centering 
\caption{
Although our formalism applies to any pair of LIMs $X$ and $Y$,
we consider the specific example of [C{\sc ii}] and Ly-$\alpha$ LIMs from redshift 5.
Although each LIM is contaminated by interlopers (CO and C{\sc i} for [C{\sc ii}], and H$\alpha$ and H$\beta$ for Ly-$\alpha$), these interlopers do not overlap in redshift, and are therefore uncorrelated.
As a result, they do not bias the LIM-pair lensing estimator, as we show below.
Neither axis is to scale in this schematic.}
\label{fig:dI_dz}
\end{figure}

While our formalism applies identically to any pair of such lines $X$ and $Y$, we focus on a specific example below.
We consider intensity maps in [C{\sc ii}] and Ly-$\alpha$ at redshift $z=5$ as our intensity maps $X$ and $Y$.
The [C{\sc ii}] LIM is contaminated by CO and C{\sc i} rotational lines from various redshifts.
Similarly, the Ly-$\alpha$ LIM is contaminated by H$\alpha$ and H$\alpha$ interlopers at low redshift.
Crucially, as illustrated in Fig.~\ref{fig:dI_dz}, the interlopers for [C{\sc ii}] and Ly-$\alpha$ do not overlap in redshift, such that they are indeed statistically independent. 
For concreteness, in what follows, we focus on CO (J=4-3) and H$\alpha$ lines as interlopers to the target [C{\sc ii}] and Ly-$\alpha$ lines respectively. Our analysis however, is equally applicable to all the interloper lines simultaneously, since they do not overlap in redshift.

A key input to the LIM-pair lensing estimator below is the auto- and cross-spectra of the LIMs $X$ and $Y$. 
Computing the effect of interlopers on the bias and variance of this estimator further requires modeling the bispectra and trispectra of these LIMs.
For all this, we use the halo model formalism from \cite{Schaan21a, Schaan21b}, based on conditional luminosity functions, and use the publicly available code 
\texttt{HaloGen}\footnote{\url{https://github.com/EmmanuelSchaan/HaloGen/tree/LIM}},
as described in App.~\ref{app:halo_model}.

\section{Line-pair lensing quadratic estimators}

To derive the LIM-pair lensing quadratic estimator, we follow Ref.~\cite{Hu02}.
We seek an estimator of the form 
\beq
\hat{\kappa}_{XY}(\bll) 
= 
\int \frac{d^2 l_1}{(2 \pi)^2}
\frac{d^2 l_2}{(2 \pi)^2} \
\delta^D_{\bl_1+\bl_2} \
F_{XY}(\bl_1, \bl_2) \ X_{\bl_1} Y_{\bL - \bl1} \ , 
\eeq
where $\bll = \bl_1 + \bl_2$ and the Dirac delta enforces the Fourier mode constraint,
and 
$F_{XY}$
is uniquely determined by requiring $\hat{\kappa}_{XY}$ to be unbiased (to first order in the true $\kappa$) and to have minimum variance.
As shown in App.~\ref{sec:derivation_lensing_qe}, the solution is
\beq
\bal
&F_{XY}(\bl_1, \bl_2) 
= \lambda_{XY}(L) \times\\
&\quad\quad\frac{C_{l_1}^{YY} C_{l_2}^{XX} f_{XY}(\bl_1, \bl_2) - C_{l_1}^{XY} C_{l_2}^{XY}  f_{XY}(\bl_2, \bl_1)}{C_{l_1}^{XX} C_{l_2}^{YY}C_{l_1}^{YY} C_{l_2}^{XX} - \left(C_{l_1}^{XY} C_{l_2}^{XY}\right)^2} \, \eal
,
\eeq
where the Lagrange multiplier $\lambda_{XY}(L)$ is given by Eq.~\eqref{eq:lagrange_xy}.
In what follows, we compare this estimator to the ones built on LIM $X$ (denoted $\hat{\kappa}_{XX}$) or $Y$  (denoted $\hat{\kappa}_{YY}$) alone, 
where $F_{XX}$ and $F_{YY}$ are given by Eq.~\eqref{eq:F_XX}.

\section{Gaussian noise bias $N^{(0)}$}

Similarly to all quadratic lensing estimators, 
the LIM-pair estimator is affected by the Gaussian lensing reconstruction noise $N^{(0)}$, given in Eq.~\ref{eq:variance}.
In particular, the lensing noise for the LIM-pair estimator $\hat{\kappa}_{XY}$
receives contribution not only from the cross-spectrum $C_\ell^{XY}$,
but also the auto-spectra $C_\ell^{XX}$ and $C_\ell^{YY}$.
Interloper foregrounds, which do not affect the cross-spectrum, do enhance the auto-spectra, thus increasing the lensing noise.
As a result, the lensing noise for $\hat{\kappa}_{XY}$
is not significantly reduced compared to those of $\hat{\kappa}_{XX}$ and $\hat{\kappa}_{YY}$.
This makes sense intuitively: although the interlopers are nulled in the cross-spectrum, they are still present in the LIMs, acting as a source of noise.

This lensing noise $N^{(0)}$ receives contribution from the power spectra of the target line itself, the detector noise, and potential foregrounds.
However, the $N^{(0)}$ noise only takes into account the Gaussian part of these components.
If the interloper foregrounds were Gaussian random fields, they would be fully described by $N^{(0)}$, and would thus be automatically subtracted by the standard $N^{(0)}$ subtraction.
Thus, they would not be a concern.
In the next subsection, we thus focus on the non-Gaussianity of interloper foregrounds, to compute their bias to LIM lensing.


\section{Non-Gaussian interloper biases can overwhelm the standard lensing estimator}

\begin{figure}[h!]
\centering
\includegraphics[width=\columnwidth]{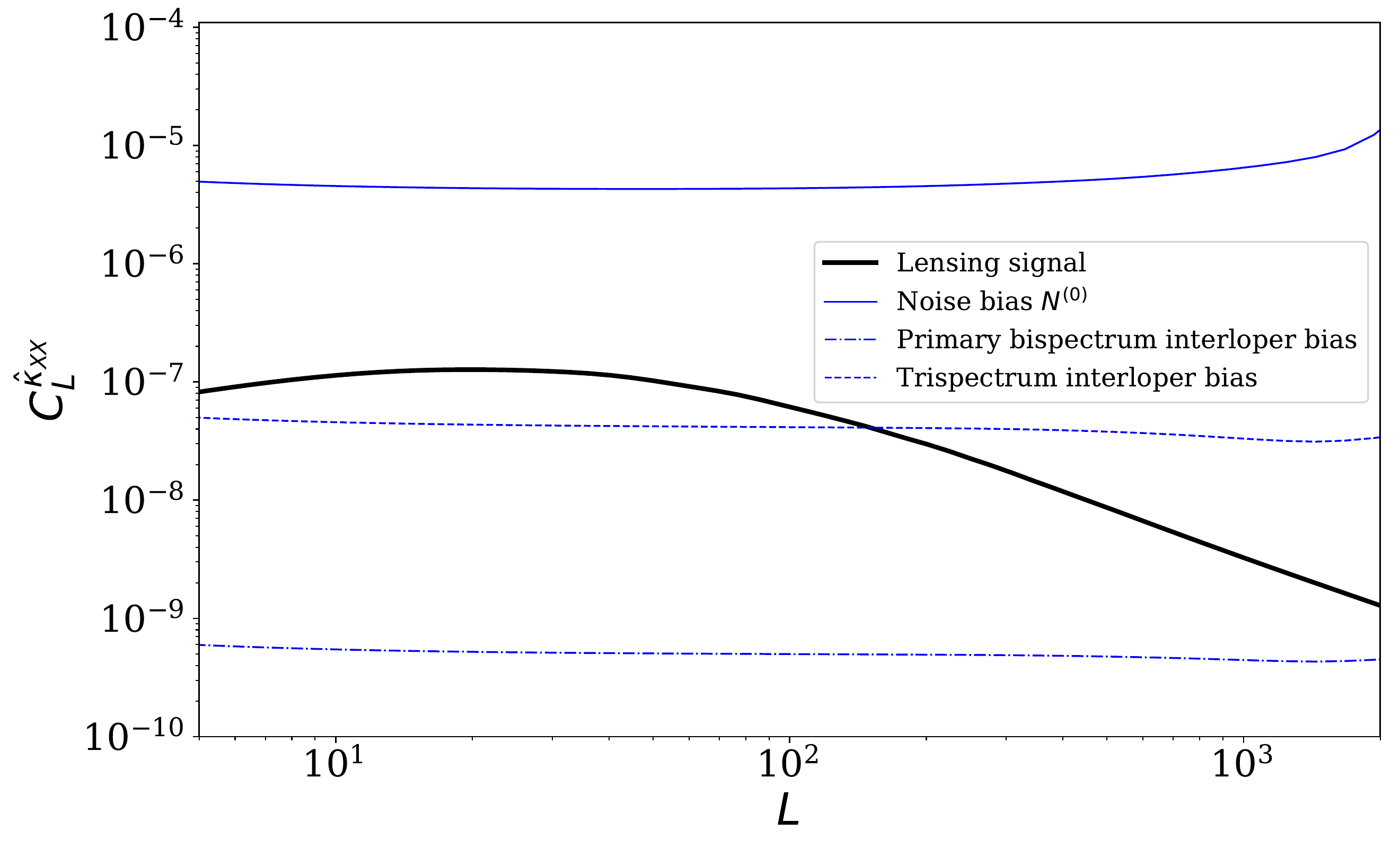}
\centering 
\caption{
For the standard LIM lensing estimator (here $\hat{\kappa}_{XX}$ with $X=$Ly-$\alpha$ at $z=5$), the lensing noise $N^{(0)}$ (light blue) is comparable to the lensing signal (solid black).
However, the interloper contamination (here H$\alpha$ at $z=0.12$) produces a dominant bias to the lensing power spectrum.
This non-Gaussian bias is the sum of the primary bispectrum (blue dot-dashed) and the trispectrum (blue dashed) terms. 
We do not show the secondary bispectrum here as it is negligible with respect to the primary bispectrum and trispectrum biases.
This motivates the need for the new LIM lensing estimator we derive in this paper.
}
\label{fig:HO02_ind}
\end{figure}

Similarly to CMB lensing, interloper foregrounds cause a bias in LIM lensing because they are non-Gaussian and correlated with the true lensing field we seek to reconstruct.
In this section, we follow the CMB lensing derivation from \cite{vanEngelen14, Osborne14, Ferraro18, Schaan19} and adapt it to the case of LIM interlopers.
We leave the detailed derivation to App.~\ref{app:biases} and instead discuss the intuitive origin of the various terms, shown in Fig.~\ref{fig:HO02_ind} for the standard (non LIM-pair) lensing estimator $\hat{\kappa}_{XX}$.

Because the lensing estimators $\hat{\kappa}$ considered here are quadratic in the LIMs, the estimated power spectrum $C_L^{\hat{\kappa}\hat{\kappa}}$ is quartic in the LIMs. 
One therefore naturally expects a bias coming from four powers of the interlopers. 
As we discussed above, the Gaussian part of this term is already included in the $N^{(0)}$ term, and therefore automatically subtracted by the $N^{(0)}$ subtraction.
Thus the remaining bias comes from the connected, non-Gaussian four point function of the interlopers, i.e. their trispectrum.
This trispectrum bias is shown with a dashed line in Fig.~\ref{fig:HO02_ind}.

Not only are the interlopers non-Gaussian, leading to the trispectrum bias above, they are also correlated with the true lensing signal we seek to reconstruct.
Indeed, the interlopers trace the large-scale mass distribution, which contributes to the true lensing of the target LIMs.
In other words, the target LIM is lensed in part by the interloper, which contaminates the observed LIM.
This effect, called ``self-lensing'' in \cite{Schaan18}, originates from the bispectrum between two powers of the interlopers and the true lensing potential.
It can be split into two terms, the so-called primary and secondary bispectrum interloper biases. 
If the two target lines contributing to the reconstructed $\kappa$ which enters the bispectrum with two interlopers belong to the pairs of multipoles $\bl_1, \bl_2$ and $\bl_3, \bl_4$ for which the lensing weights $F_{XX}(\bl_1, \bl_2)$ and $F_{XX}(\bl_3, \bl_4)$ optimize the quadratic estimator, we get the primary bispectrum bias. If that is not the case e.g. if one target line comes from $\bl_1$ and the other one from $\bl_3$ with lensing weights $F_{XX}(\bl_1, \bl_2)$ and $F_{XX}(\bl_3, \bl_4)$, this gives rise to an inefficient $\kappa$ reconstruction which enters the bispectrum and thus is called the secondary bispectrum. This is discussed in detail in App.~\ref{app:biases}.
In this analysis, we consider only the 1-halo term of the trispectrum and bispectrum biases, giving a lower bound to the total interloper bias.
We find the primary bispectrum to be smaller than the lensing signal (dot-dashed line in Fig.~\ref{fig:HO02_ind}), and that the secondary bispectrum is negligible.
However, the trispectrum bias term (dotted line in Fig.~\ref{fig:HO02_ind}) for $\hat{\kappa}_{XX}$ is comparable to the lensing signal for $L\lesssim 200$ and dominant for higher lensing multipoles.
In consequence, the standard LIM lensing reconstruction method is highly biased by interlopers, and another method is needed to control them.

\section{Avoiding all biases with the LIM-pair $\times$ CMB lensing cross-spectrum}

\subsection{Avoiding all interloper biases with LIM-only lensing?}

We have shown that the lensing power spectrum estimated from 
$\hat{\kappa}_{XX}\hat{\kappa}_{XX}$
is biased by the primary, secondary and trispectrum terms.
We may instead try to use different combinations of the $X$
and $Y$ LIMs, to reconstruct the lensing power spectrum.

\begin{figure}[h!]
\centering
\includegraphics[width=\columnwidth]{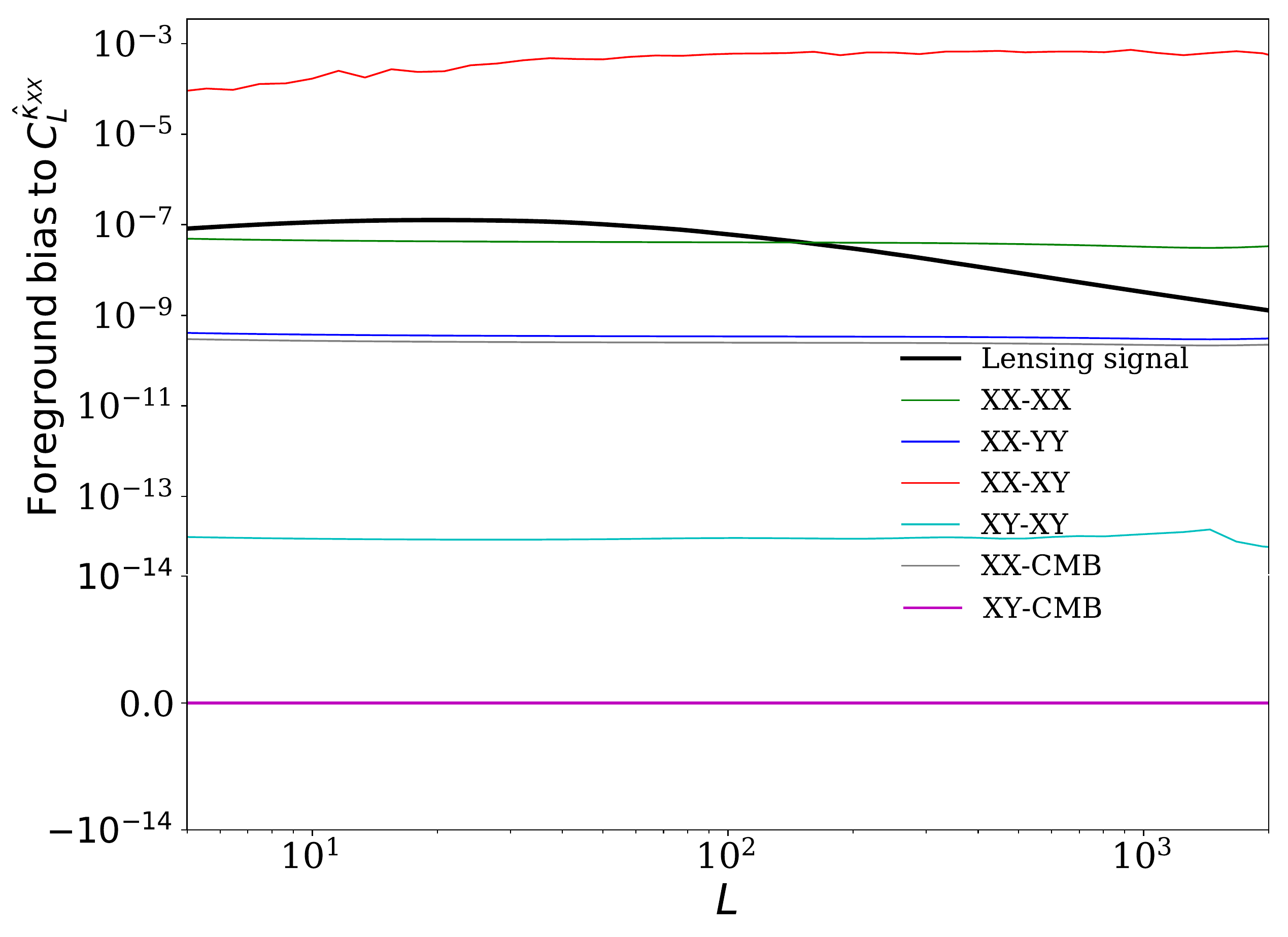}
\centering 
\caption{
Even with two LIMs $X$=Ly-$\alpha$ and $Y$=[C{\sc ii}] at $z=5$, whose interlopers are independent, one cannot avoid all the interloper biases.
The combinations 
$\hat{\kappa}_{XX}\hat{\kappa}_{XX}$ (green),
$\hat{\kappa}_{XX}\hat{\kappa}_{XY}$ (red)
and
$\hat{\kappa}_{XY}\hat{\kappa}_{XY}$ (cyan)
are dominated by the residual secondary bispectrum term.
The combinations 
$\hat{\kappa}_{XX}\hat{\kappa}_{YY}$ (blue)
and
$\hat{\kappa}_{XX}\hat{\kappa}_\text{CMB}$ (grey)
are dominated by the residual primary bias.
However, the cross-correlation of the LIM-pair estimator and CMB lensing,
i.e.
$\hat{\kappa}_{XY}\hat{\kappa}_\text{CMB}$ (purple)
is entirely free of interloper bias.
This is the main result of this paper.}
\label{fig:biases}
\end{figure}

The combination
$\hat{\kappa}_{XX}\hat{\kappa}_{YY}$
avoids the interloper trispectrum, since the interlopers in $X$ and $Y$ originate from different redshifts, and are therefore independent.
This combination also avoids the secondary bispectrum bias.
However, it is not free of primary bispectrum bias, making it still largely biased by interlopers.
The combination
$\hat{\kappa}_{XX}\hat{\kappa}_{XY}$
is free of trispectrum bias, but not of primary or secondary bispectrum biases.
For this lensing cross-spectrum, the interloper bias is dominant and comes mostly from the secondary bispectrum.
Finally, the combination 
$\hat{\kappa}_{XY}\hat{\kappa}_{XY}$
avoids the trispectrum and primary bispectrum terms,
but still suffers from the secondary bispectrum bias. 
However, we find that the secondary bias for $\hat{\kappa}_{XY}\hat{\kappa}_{XY}$ is small and can potentially be neglected. 
In short, combinations from two LIMs $X$ and $Y$ cannot suppress all the interloper bias terms,
but the auto-spectrum of the ``LIM-pair'' lensing estimator appears to sufficiently reduce them.
While the bias to $\hat{\kappa}_{XY}\hat{\kappa}_{XY}$ appears negligible (secondary bispectrum only), our bispectrum calculation only includes the 1-halo term, such that it is only a lower limit.
Furthermore, the secondary bias may be larger when considering different pairs of lines.
The interloper biases for the various combinations are shown in Fig.~\ref{fig:biases}.
Interestingly, in Fig.~\ref{fig:biases}, the interloper bias to lensing is very different for $\hat{\kappa}_{XX}\hat{\kappa}_{XY}$ and $\hat{\kappa}_{XY}\hat{\kappa}_{XY}$, even though they are both dominated by secondary bispectrum-like terms.
We explain this in App.~\ref{app:biases}.

Using three LIMs $X$, $Y$ and $Z$ from the same redshift, with independent interlopers, still does not avoid all the interloper biases.
If four LIMs $X$, $Y$, $Z$ and $W$ were available from the same redshift, with independent interlopers, the combination 
$\hat{\kappa}_{XY}\hat{\kappa}_{ZW}$
would be entirely free of interloper bias.
Although one may hope to use CO, [C{\sc ii}], Ly-$\alpha$ and 21~cm LIMs from the same redshift, this prospect remains futuristic.

\subsection{Avoiding all the biases via CMB lensing cross-correlation}

In order to further suppress interloper biases, we now turn to cross-correlations of LIM-lensing with CMB lensing.
The combination
$\hat{\kappa}_{XX} \hat{\kappa}_\text{CMB}$
is free of trispectrum and secondary bispectrum bias,
but it still suffers from the primary bispectrum.
As a result, it does not reduce the interloper bias, as illustrated in Fig.~\ref{fig:biases}.

On the other hand, the combination 
$\hat{\kappa}_{XY} \hat{\kappa}_\text{CMB}$
is entirely free of interloper biases: it is not affected by the primary and secondary bispectra, nor the trispectrum.
This is the main result of this paper:
LIM lensing can be measured without any interloper bias, when cross-correlating the LIM-pair estimator with CMB lensing.
Given the uncertain and potentially large interloper biases for the standard LIM lensing estimators, this constitutes a dramatic progress.

\subsection{Detectability: Signal-to-noise ratio}

\begin{figure}
\centering
\includegraphics[width=\columnwidth]{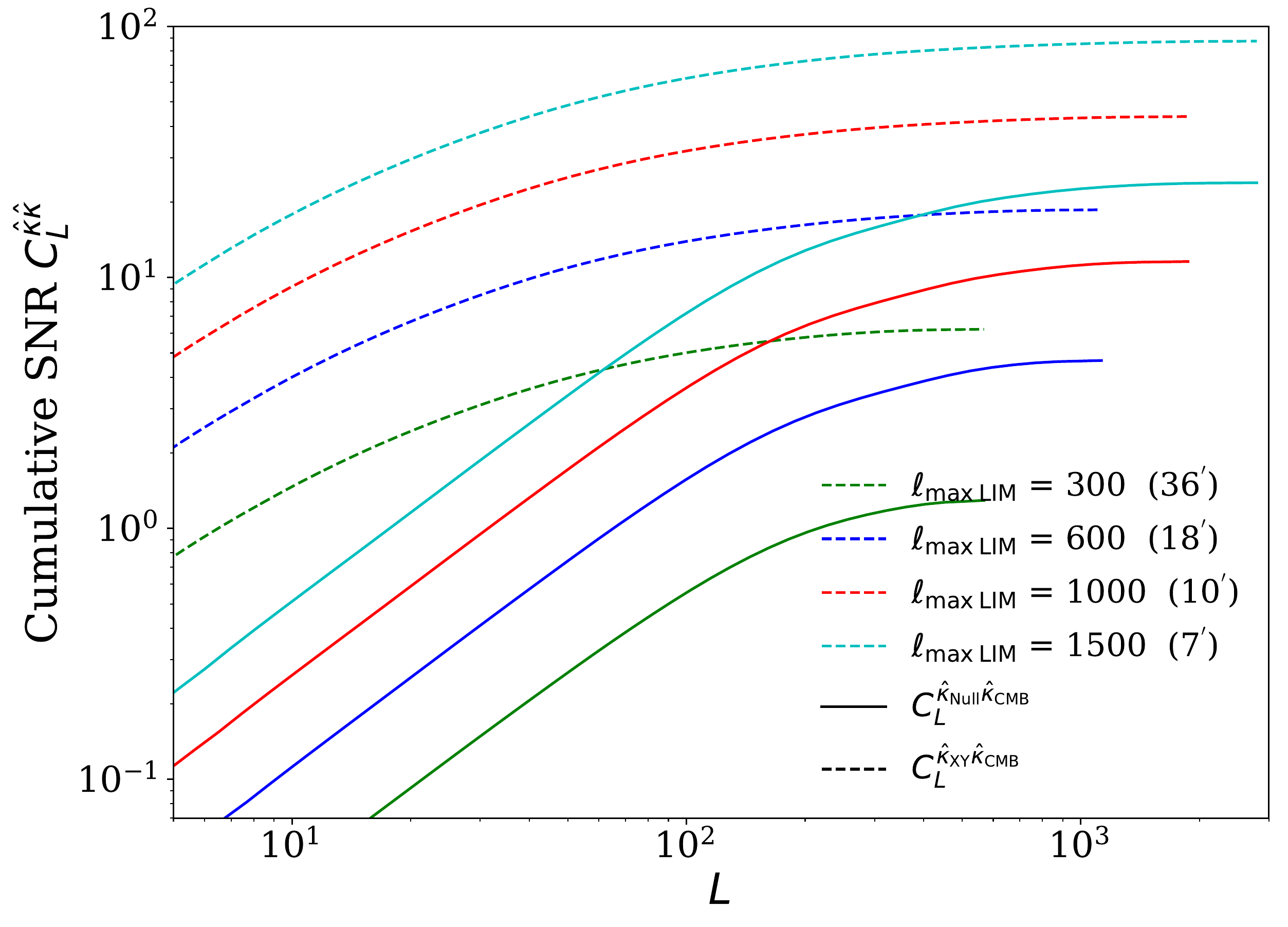}
\centering 
\caption{
Including Gaussian noise and the noise from the non-Gaussian interlopers, cumulative SNR for the LIM $\times$ CMB lensing cross-power spectrum where $X=$~Ly-$\alpha$ and $Y=$~[C{\sc ii}] LIM are shown in dashed lines for different $\ell_\text{max LIM}$. 
Solid lines show the corresponding cumulative SNR for the "nulled" $\hat{\kappa}_\text{Null}$ $\times$ CMB lensing power spectrum having contributions only from $z > 5$. 
Both the power spectra are detectable for a simple idealized experiment where the LIMs are signal dominated over detector noise out to $\ell_\text{max LIM}\sim 1000$ at $z=5$. 
The SNR is calculated with $f_{\rm sky} = 0.4$, and the CMB lensing is assumed to be noiseless out to $L=1500$ as appropriate for Simons Observatory (SO) \cite{Ade_19}.
For different $\ell_\text{LIM max}$ values, we provide the minimum angular scale the beam will have to resolve (calculated simply as $180^\circ/\ell_\text{LIM max}$) in arcminutes.
}
\label{fig:cumsnr_difflmax}
\end{figure}

In this section, we answer the question of the detectability of the 
$C_L^{\hat{\kappa}_\text{LIM} \hat{\kappa}_\text{CMB}}$ and $C_L^{\hat{\kappa}_\text{null} \hat{\kappa}_\text{CMB}}$ i.e. the
cross-spectrum of the CMB lensing with LIM-pair estimator and the "nulled" estimator respectively by computing its expected SNR.
We consider an idealized and futuristic experiment, signal-dominated in the LIMs out to $\ell_\text{max LIM} = 300 - 1500$.
Our SNR calculation is described in detail in App.~\ref{app:snr_lensing_cross}.
While it is technically an upper limit, we expect it to also be a good approximation to the truth.
In short, we adopt the Gaussian SNR formula, including the lensing noise $N^{(0)}$ as well as the non-Gaussian terms $\mathcal{B}^p$, $\mathcal{B}^s$, and $\mathcal{T}$ from interlopers in the noise for $C_L^{\hat{\kappa}_\text{LIM} \hat{\kappa}_\text{LIM}}$. 
As $\hat{\kappa}_\text{Null}$ is constructed through a combination of $\hat{\kappa}_{XY}$ and $\hat{\kappa}_{\rm CMB}$, the $XY$ part adds a secondary bispectrum bias which as we show in Fig.~\ref{fig:biases} is quite small and can be neglected here. Thus we consider only the $N^{(0)}$ terms for $C_L^{\hat{\kappa}_\text{Null} \hat{\kappa}_\text{CMB}}$ SNR calculation.
The various angular resolutions assumed are conservative for the lines we consider (Ly-$\alpha$ and [C{\sc ii}]).
For instance, an experiment like CONCERTO \cite{Concerto20} should measure the [C{\sc ii}] line at $z=5$ with $0.24'$ resolution, significantly higher than assumed here.
SPHEREx \cite{Dore14, Dore18} is expected to produce a Ly-$\alpha$ LIM at $z=5$ with $6''$ resolution, even much higher.
As Fig.~\ref{fig:cumsnr_difflmax} shows, the SNR on $C_L^{\hat{\kappa}_\text{LIM} \hat{\kappa}_\text{CMB}}$ may reach several 10s of~$\sigma$, allowing for a significant detection of the LIM $\times$ CMB lensing cross-power spectrum. 
At the same time the SNR for $C_L^{\hat{\kappa}_\text{Null} \hat{\kappa}_\text{CMB}}$ is slightly lower which is expected but it may still be significantly detected with an experiment like we have considered here. For the detector noise, an experiment with sensitivity like CONCERTO over a large sky fraction will be required for such a detection whereas the sensitivity of a SPHEREx like experiment may not be sufficient. 
As for any LIM forecast, the theoretical uncertainty on the LIM power spectra at high redshift is very large, which may affect our conclusions.
We relied on the halo model predictions from \cite{Schaan21a, Schaan21b}, whose LIM power spectra were found in agreement with the literature. 

While upcoming experiments may be limited by sensitivity and sky coverage, a futuristic experiment such as the one we considered here can thus detect LIM lensing with the LIM-pair lensing and the null combination, in cross-correlation with CMB lensing.
This therefore offers a powerful way to probe the high redshift Universe.

\section{Conclusion}

Lensing from LIMs has the potential to allow lensing tomography at higher redshift than galaxy surveys, and to provide a new probe of the high-redshift Universe.
We show that the nulling technique allows us to selectively extract the matter density field at $z=1-5$ in combination with galaxy lensing, and at $z>5$ in combination with CMB lensing,

However, interloper foregrounds contaminating LIMs are a major hurdle to LIM lensing. 
In this paper, we quantified the lensing bias from interlopers for the first time, showing it to be very significant for the standard LIM lensing estimators.
We derived a new LIM-pair lensing estimator, based on two LIMs in different lines, from the same redshift, with independent interlopers.
In cross-correlation with CMB lensing, it exactly nulls all the interloper bias terms, which would otherwise dominate.

When using the standard lensing estimator, the non-Gaussian interlopers can also largely enhance the lensing noise.
This enhancement is uncertain because it depends on our modeling of LIM bispectra and trispectra.
In contrast, the LIM-pair lensing estimator, in cross-correlation with CMB lensing, is exactly free of interloper bias, insensitive to these modeling uncertainty, making it dramatically more reliable.

We have shown that a simple, idealized LIM experiment can detect LIM lensing at $z=5$, provided that the detector noise is subdominant to the target lines lines in the pair estimator (here Ly-$\alpha$ and [C{\sc ii}]). 

We have not addressed the biases to LIM lensing from the non-Gaussianity of the target lines, rather than their interlopers.
These were studied in \cite{Schaan18, Foreman18} and a bias-hardened estimator was derived to control these biases \cite{Foreman18}.
We have also not addressed the LIM lensing biases from continuum foregrounds, and assumed that they can be controlled by discarding the low $k_\parallel$ modes in the LIMs.
Finally, we have not quantified the bias due to the fact that the interlopers are themselves lensed.
Similarly to the case of CMB lensing \cite{Mishra19}, we expect this bias to be small. 
Combining the interloper removal techniques like voxel masking \cite{Kovetz17, Pullen13} with the LIM-pair estimator will further help and we leave this study for future work.
If the future studies improve upon the interloper cleaning in the LIMs, the quadratic estimator which we propose here could potentially detect $C_L^{\hat{\kappa}_\text{LIM} \hat{\kappa}_\text{CMB}}$ with even higher SNR.

\acknowledgments

We thank Yacine Ali-Ha\"imoud, Patrick Breysse, Yun-Ting Cheng, Simone
Ferraro, Simon Foreman, Adam Lidz, Adrian Liu and Martin White for their helpful feedback on an early version of the manuscript. 
E.S. thanks Francis Bernardeau for a helpful discussion of nulling in the context of galaxy lensing, and Simone Ferraro for helpful discussions on the sensitivity of CMB lensing to very early matter density fluctuations.
E.S. is supported by the Chamberlain fellowship at Lawrence Berkeley National Laboratory.  A.R.P. was supported by NASA under award numbers 80NSSC18K1014 and NNH17ZDA001N.

\bibliographystyle{prsty.bst}
\bibliography{refs}

\begin{thebibliography}{10}

\bibitem{Planck18I}
{Planck Collaboration} {\it et~al.}, \aap {\bf 641},  A1  (2020).

\bibitem{Kovetz17}
E.~D. {Kovetz} {\it et~al.}, arXiv e-prints  arXiv:1709.09066  (2017).

\bibitem{Haslam82}
C.~G.~T. {Haslam}, C.~J. {Salter}, H. {Stoffel}, and W.~E. {Wilson}, \aaps {\bf
  47},  1  (1982).

\bibitem{Rengelink97}
R.~B. {Rengelink} {\it et~al.}, \aaps {\bf 124},  259  (1997).

\bibitem{Santos05}
M.~G. {Santos}, A. {Cooray}, and L. {Knox}, \apj {\bf 625},  575  (2005).

\bibitem{Lewis06}
A. {Lewis} and A. {Challinor}, \physrep {\bf 429},  1  (2006).

\bibitem{SO19}
P. {Ade} {\it et~al.}, \jcap {\bf 2019},  056  (2019).

\bibitem{CMBS419}
K. {Abazajian} {\it et~al.}, arXiv e-prints  arXiv:1907.04473  (2019).

\bibitem{McCarthy21}
F. McCarthy, S. Foreman, and A. van Engelen, Phys. Rev. D {\bf 103},  103538
  (2021).

\bibitem{Zahn06}
O. {Zahn} and M. {Zaldarriaga}, \apj {\bf 653},  922  (2006).

\bibitem{Pourtsidou14}
A. {Pourtsidou} and R.~B. {Metcalf}, \mnras {\bf 439},  L36  (2014).

\bibitem{Pourtsidou15}
A. {Pourtsidou} and R.~B. {Metcalf}, \mnras {\bf 448},  2368  (2015).

\bibitem{Pourtsidou16}
A. {Pourtsidou}, D. {Bacon}, R. {Crittenden}, and R.~B. {Metcalf}, \mnras {\bf
  459},  863  (2016).

\bibitem{Schaan18}
E. {Schaan}, S. {Ferraro}, and D.~N. {Spergel}, \prd {\bf 97},  123539  (2018).

\bibitem{Foreman18}
S. {Foreman}, P.~D. {Meerburg}, A. {van Engelen}, and J. {Meyers}, \jcap {\bf
  2018},  046  (2018).

\bibitem{Chakraborty19}
P. {Chakraborty} and A.~R. {Pullen}, \mnras {\bf 488},  1828  (2019).

\bibitem{Feng19}
C. {Feng} and G. {Holder}, arXiv e-prints  arXiv:1905.02084  (2019).

\bibitem{LSSTScienceBook}
{LSST Science Collaboration} {\it et~al.}, ArXiv e-prints  (2009).

\bibitem{Doux16}
C. {Doux} {\it et~al.}, \prd {\bf 94},  103506  (2016).

\bibitem{Huterer05}
D. {Huterer} and M. {White}, \prd {\bf 72},  043002  (2005).

\bibitem{Bernardeau14}
F. {Bernardeau}, T. {Nishimichi}, and A. {Taruya}, \mnras {\bf 445},  1526
  (2014).

\bibitem{Barthelemy20}
A. {Barthelemy} {\it et~al.}, \mnras {\bf 492},  3420  (2020).

\bibitem{Zhu18}
H.-M. {Zhu}, U.-L. {Pen}, Y. {Yu}, and X. {Chen}, \prd {\bf 98},  043511
  (2018).

\bibitem{Osborne14}
S.~J. {Osborne}, D. {Hanson}, and O. {Dor{\'e}}, \jcap {\bf 2014},  024
  (2014).

\bibitem{Namikawa13}
T. {Namikawa}, D. {Hanson}, and R. {Takahashi}, \mnras {\bf 431},  609  (2013).

\bibitem{Planck13XVII}
P.~A.~R. Ade {\it et~al.}, Astronomy \& Astrophysics {\bf 571},  A17  (2014).

\bibitem{Sailer20}
N. {Sailer}, E. {Schaan}, and S. {Ferraro}, \prd {\bf 102},  063517  (2020).

\bibitem{Pullen13}
A.~R. {Pullen}, O. {Dor{\'e}}, and J. {Bock}, \apj {\bf 786},  111  (2014).

\bibitem{Gong14}
Y. {Gong}, M. {Silva}, A. {Cooray}, and M.~G. {Santos}, \apj {\bf 785},  72
  (2014).

\bibitem{Breysee15}
P.~C. {Breysse}, E.~D. {Kovetz}, and M. {Kamionkowski}, \mnras {\bf 452},  3408
   (2015).

\bibitem{Yue15}
B. {Yue} {\it et~al.}, \mnras {\bf 450},  3829  (2015).

\bibitem{Silva15}
M. {Silva}, M.~G. {Santos}, A. {Cooray}, and Y. {Gong}, \apj {\bf 806},  209
  (2015).

\bibitem{Sun18}
G. {Sun} {\it et~al.}, \apj {\bf 856},  107  (2018).

\bibitem{Cheng20}
Y.-T. {Cheng}, T.-C. {Chang}, and J.~J. {Bock}, \apj {\bf 901},  142  (2020).

\bibitem{Visbal10}
E. {Visbal} and A. {Loeb}, \jcap {\bf 2010},  016  (2010).

\bibitem{Cheng16}
Y.-T. {Cheng} {\it et~al.}, \apj {\bf 832},  165  (2016).

\bibitem{Liu16}
A. {Liu}, Y. {Zhang}, and A.~R. {Parsons}, \apj {\bf 833},  242  (2016).

\bibitem{Lidz16}
A. {Lidz} and J. {Taylor}, \apj {\bf 825},  143  (2016).

\bibitem{Gong20}
Y. {Gong}, X. {Chen}, and A. {Cooray}, arXiv e-prints  arXiv:2001.10792
  (2020).

\bibitem{Madhavacheril18}
M.~S. {Madhavacheril} and J.~C. {Hill}, \prd {\bf 98},  023534  (2018).

\bibitem{Darwish21}
O. {Darwish} {\it et~al.}, \mnras {\bf 500},  2250  (2021).

\bibitem{Schaan21a}
E. {Schaan} and M. {White}, arXiv e-prints  arXiv:2103.01964  (2021).

\bibitem{Schaan21b}
E. {Schaan} and M. {White}, arXiv e-prints  arXiv:2103.01971  (2021).

\bibitem{Hu02}
W. {Hu} and T. {Okamoto}, \apj {\bf 574},  566  (2002).

\bibitem{vanEngelen14}
A. {van Engelen} {\it et~al.}, \apj {\bf 786},  13  (2014).

\bibitem{Ferraro18}
S. {Ferraro} and J.~C. {Hill}, \prd {\bf 97},  023512  (2018).

\bibitem{Schaan19}
E. {Schaan} and S. {Ferraro}, \prl {\bf 122},  181301  (2019).

\bibitem{Concerto20}
{The CONCERTO collaboration} {\it et~al.}, arXiv e-prints  arXiv:2007.14246
  (2020).

\bibitem{Dore14}
O. {Dor{\'e}} {\it et~al.}, arXiv e-prints  arXiv:1412.4872  (2014).

\bibitem{Dore18}
O. {Dor{\'e}} {\it et~al.}, arXiv e-prints  arXiv:1805.05489  (2018).

\bibitem{Ade_19}
P. {Ade} {\it et~al.}, \jcap {\bf 2019},  056  (2019).

\bibitem{Mishra19}
N. {Mishra} and E. {Schaan}, \prd {\bf 100},  123504  (2019).

\bibitem{Maniyar21}
A.~S. {Maniyar} {\it et~al.}, \prd {\bf 103},  083524  (2021).

\bibitem{Bohm20}
V. {B{\"o}hm}, C. {Modi}, and E. {Castorina}, \jcap {\bf 2020},  045  (2020).

\end{thebibliography}

\newpage
\onecolumngrid
\appendix

\section{Halo models for the lines considered}
\label{app:halo_model}

Throughout the paper, the LIM auto- and cross-spectra are computed following \cite{Schaan21a, Schaan21b}, using the public code \texttt{HaloGen}\footnote{\url{https://github.com/EmmanuelSchaan/HaloGen/tree/LIM}}.
We take the Fourier space 3D power spectra, bispectra, and trispectra from \texttt{HaloGen} and convert them into their respective angular space 2D forms using the thin shell approximation. Working within the Limber approximation, for the power spectrum this conversion is done as
\beq
C_\ell \approx \mathcal{V}^{-1} P(k=\ell/\chi_0, z) \, ,
\eeq
where $\mathcal{V} = \chi^2_0 \Delta\chi$ is the volume per steradian.

To compute the higher-point functions, specifically bispectra and trispectra, we extend this code as follows.
For the bispectrum between interloper lines and the CMB lensing convergence, we only evaluate the 1-halo term:
\beq
\bal
B^{\kappa II, 1h}(\bl_1, \bl_2, \bl_3, z)
&=
\int \frac{d\chi}{\chi^4}\ 
W_\kappa(\chi)
W_I(\chi)^2
\left(\frac{c}{4\pi \nu_g^0 H(z)}\right)^2 
&\int dm \; n(m)\
\frac{m}{\bar{\rho}}
L_g(m)^2\
u_m(\frac{\ell_1}{\chi},m)
u_g(\frac{\ell_2}{\chi},m)
u_g(\frac{\ell_3}{\chi},m)
,
\eal
\label{eq:b1h}
\eeq
where $W_\kappa$ is the lensing kernel of interest (e.g., CMB or LIM lensing),
and $W_I(\chi)$ simply describes the redshift distribution of the interloper line emitters,
i.e.
$W_I(\chi) = \mathbb{I}_{\chi \in [\chi_0, \chi_0+\Delta\chi]} / \Delta \chi$.
To speed up the multiple integrals, we shall further approximate
$u_m(k_1,m)
\sim u_A(k_2,m)
\sim u_B(k_3,m)
\sim 1$
on the scales considered,
such that $B^{\kappa A B, 1h}(z)$ only needs to be evaluated once per redshift.\\
For the trispectrum, we only evaluate the one halo term, resulting in a lower limit to the trispectrum.
\beq
\mathcal{T}^{g\ \text{shot}}(z)
=
\int \frac{d\chi}{\chi^6}\
W_g(\chi)^4\
\left(\frac{c}{4\pi \nu_g^0 H(z)}\right)^4 
\int dm \; n(m)\
\int dL_g\ \kappa(L_g|m) L_g^4.
\eeq

It has to be noted that for the Ly-$\alpha$ line we consider at $z=5$, the H$\alpha$ line at $z\approx0.12$ acts as an interloper. Thus all the power spectrum, bispectrum, and trispectrum corresponding to H$\alpha$ line have to evaluated at $z=0.12$. \texttt{HaloGen} code we use for this purpose, relies on the observed luminosity functions of the galaxies which are only available in certain redshift ranges. In case of H$\alpha$ line, unfortunately this is not available at $z \approx 0.1$. In this case, we calculate the approximated power spectra and other moments by assuming that the ratio of these moments at two different redshifts where the luminosity functions are available vary linearly with the ratio of the two redshifts. For example, we first calculate the power spectrum of H$\alpha$ line at $z=0.8$ and $z=0.4$, and assume that the change in power spectrum at these two redshifts scales the same way to $z=0.12$ and then obtain the power spectrum at $z=0.12$. This procedure will not give us the true power spectrum and other moments, however, that does not affect the results of our work.

\section{LIM-pair lensing quadratic estimator: extending HO02}
\label{sec:derivation_lensing_qe}

\cite{Maniyar21} present a discussion on the HO02 and slightly modified versions of HO02 quadratic estimators used in various CMB lensing analysis till date. We will build up on the HO02 estimator. Here we will work in the flat-sky approximation. $\bl$ are the two-dimensional Fourier wavenumbers for LIM and $\bll$ for the lensing potential.

The power spectra of the observed LIM fields are defined as
\barr
\langle X(\bl) Y(\bl') \rangle = (2\pi)^2 \delta(\bl+\bl') C_l^{XY} \, ,
\label{eq:limps}
\earr
where $C_l^{XY}$ is the total cross-power spectrum between the Gaussian lensed fields. It can also include contributions from other sources of variance such as residual foreground contamination or interlopers from different redshifts. The angular brackets here denote taking ensemble averages over the primordial CMB, along with the underlying large scale structure.

The observed LIM are lensed due to the matter distribution between the redshift at which the target line was emitted and us. This lensing of the LIM results in different Fourier modes of a given map being correlated with each other which would not be the case for a Gaussian unlensed field. Using these correlations, if we can model the power spectrum of the un-lensed LIM, we can reconstruct the lensing potential $\kappa$.
\beq
  \left\langle \frac{\delta}{\delta \kappa(\bll)}\left(  X(\bl)Y(\bl')\right)\right\rangle
  = \delta(\bl+\bl'-\bll) f_{XY}(\bl,\bl').
\label{eq:limavg}
  \eeq
where $f_{XY}(\bl, \bl')$ is the coupling coefficient given as
\beq
f_{XY}(\bl, \bl') = 
-\frac{2}{L^2}
\left[
\widetilde{C}_{l_1}^{XY}  \dotfac{1} 
+  \widetilde{C}_{l_2}^{XY}  \dotfac{2} 
\right]
\ ,
\label{eq:f_xy}
\eeq
where $\widetilde{C}_{l}^{XY}$ is the unlensed cross-power spectrum. It is to be noted that Eq.~\ref{eq:limps}-\ref{eq:f_xy} are applicable for a single LIM as well i.e. if $X=Y$.

For brevity, we introduce the compact notation
\beq
\intL ... \equiv \iint \frac{d^2 l_1 d^2 l_2}{(2 \pi)^2} \delta(\bl_1 + \bl_2 - \bll) ...
\eeq

From Eq.~\ref{eq:f_xy}, we can see that using suitable weights over pairs of Fourier modes, it is possible to reconstruct the $\kappa$ field:
\beq
\hat{\kappa}_{XY}(\bll) = \intL X(\bl_1) Y(\bl_2) F_{XY}(\bl_1, \bl_2) \, , 
\eeq
where the weights $F_{XY}(\bl_1, \bl_2)$ have to be determined to minimize the variance of the estimator under the constraint
\beq
\intL f_{XY}(\bl_1, \bl_2) F_{XY}(\bl_1, \bl_2) = 1 \, ,
\label{eq:unbiasedqe}
\eeq
which ensures that the estimator is unbiased, to first order in $\kappa$. 

The variance (or reconstruction noise) $N_{XY}$ is given as
\beq
\langle \hat{\kappa}_{XY}(\bll) \hat{\kappa}_{XY}(\bll') \rangle = (2\pi)^2 \delta(\bll+\bll') N_{XY}(L). \label{eq:est-noise}
\eeq
In general, Eq.~\ref{eq:est-noise} becomes
\barr
N_{XY}(L)
&=& 
\intL F_{XY}(\bl_1, \bl_2) \Big( F_{XY}(\bl_1, \bl_2) C_{l_1}^{XX} C_{l_2}^{YY} + F_{XY}(\bl_2, \bl_1) C_{l_1}^{XY} C_{l_2}^{XY} \Big).~~~ \label{eq:variance}
\earr
Minimizing this variance under the constraint \eqref{eq:unbiasedqe} results in 
\barr
\label{eq:F_XY}
F_{XY}(\bl_1, \bl_2) &=& \lambda_{XY}(L)\frac{C_{l_1}^{YY} C_{l_2}^{XX} f_{XY}(\bl_1, \bl_2) - C_{l_1}^{XY} C_{l_2}^{XY}  f_{XY}(\bl_2, \bl_1)}{C_{l_1}^{XX} C_{l_2}^{YY}C_{l_1}^{YY} C_{l_2}^{XX} - \left(C_{l_1}^{XY} C_{l_2}^{XY}\right)^2}, \\
\lambda_{XY}(L) &\equiv& \Bigg[\intL f_{XY}(\bl_1, \bl_2) \frac{C_{l_1}^{YY} C_{l_2}^{XX} f_{XY}(\bl_1,\bl_2) - C_{l_1}^{XY} C_{l_2}^{XY} f_{XY}(\bl_2,\bl_1)}{C_{l_1}^{XX} C_{l_2}^{YY}C_{l_1}^{YY} C_{l_2}^{XX} - \left(C_{l_1}^{XY} C_{l_2}^{XY}\right)^2} \Bigg]^{-1} \, .
\label{eq:lagrange_xy}
\earr

This estimator is similar to the Hu and Okamoto 2002 (HO02) estimator for CMB lensing.

Applying this estimator on LIM $X$ alone, we get
\barr
\label{eq:F_XX}
F_{XX}(\bl_1, \bl_2) &=& \lambda_{XX}(L) ~\frac{f_{XX}(\bl_1, \bl_2)}{C_{l_1}^{XX} C_{l_2}^{XX}}, \\
\lambda_{XX}(L) &\equiv& \left[\intL \frac{[f_{XX}(\bl_1, \bl_2)]^2}{C_{l_1}^{XX} C_{l_2}^{XX}}\right]^{-1}.
\earr

\section{Non-Gaussian interloper biases}
\label{app:biases}

Let us calculate the primary- and secondary-bispectrum bias terms for the lensing power spectrum. Here, we will present the calculation for the $XX$ quadratic estimator. 
\beq
X(\bl_1) = t(\bl_1) + g(\bl_1) \, ,
\eeq
where $t$ and $g$ denote the target line and the corresponding interloper line respectively in the map X. 
With this, we have
\beq
\langle \hat{\kappa}_{XX}(\bll) \hat{\kappa}_{XX}(\bll') \rangle = \intL \intLp F_{XX}(\bl_1, \bl_2) F_{XX}(\bl_3, \bl_4) \langle X(\bl_1) X(\bl_2) X(\bl_3) X(\bl_4) \rangle (2\pi)^2 \delta(\bll+\bll')
\label{eq:bispec_ensemble}
\eeq 
The ensemble average in Eq.~\ref{eq:bispec_ensemble} $\langle X(\bl_1) X(\bl_2) X(\bl_3) X(\bl_4) \rangle$ can be decomposed and summarized as follows.
\begin{table}[h]
\begin{tabular}{|l|c|}
\hline
Target signal $\langle \kappa \kappa \rangle$ & $\big\langle \big(t(\bl_1) t(\bl_2)\big) \big(t(\bl_3) t(\bl_4)\big) \big\rangle_c$ 
\vsp
Primary bispectrum $\mathcal{B}^{\kappa g g}$ & $\big\langle \big(t(\bl_1) t(\bl_2)\big) \big(g(\bl_3) g(\bl_4)\big) \big\rangle_c$ + $\big\langle \big(g(\bl_1) g(\bl_2)\big) \big(t(\bl_3) t(\bl_4)\big) \big\rangle_c$
\vsp
Secondary bispectrum $\mathcal{B}^{\kappa g g}$ & $\big\langle \big(t(\bl_1) g(\bl_2)\big) \big(t(\bl_3) g(\bl_4)\big) \big\rangle_c$ + 3 permutations
\vsp
Trispectrum $\mathcal{T}^{g g g g}$ & $\big\langle \big(g(\bl_1) g(\bl_2)\big) \big(g(\bl_3) g(\bl_4)\big) \big\rangle_c$   
\vsp
\hline
\end{tabular}
\centering \caption{Different terms in the expansion of the ensemble average of Eq.~\ref{eq:bispec_ensemble}. A detailed description of each term is given in the text.}
\label{tab:all_terms}
\end{table}

After combining factors outside of the ensemble average in Eq.~\ref{eq:bispec_ensemble} with terms in Tab.~\ref{tab:all_terms}, we get
\begin{itemize}
  \item \textbf{Target signal:} the first term gives the desired $\kappa$ power spectrum $C_L^{\hat{\kappa} \hat{\kappa}}$ after removing the Gaussian noise bias
  \item \textbf{Primary bispectrum bias to $C_L^{\hat{\kappa} \hat{\kappa}}$:} this bispectrum comes from the correlation between the $\kappa$ and foreground interloper $g$ at the same redshift. 
  This term becomes
  \barr
  \langle \hat{\kappa}_{XX}(\bll) \hat{\kappa}_{XX}(\bll') \rangle_B &=& \Bigg[ \intLp F_{XX}(\bl_3, \bl_4) \mathcal{B}_{\bll, \bl_3, \bl_4}^{\kappa g g} + \intL F_{XX}(\bl_1, \bl_2) \mathcal{B}_{\bll', \bl_1, \bl_2}^{\kappa g g} \Bigg] (2\pi)^2 \delta(\bll + \bll') \nonumber \\
  &=& 2 \intL F_{XX}(\bl_1, \bl_2) \mathcal{B}_{-\bll, \bl_1, \bl_2}^{\kappa g g} \, ,
  \label{eq:prim_bispec}
  \earr
  since the two integrals are equal.
  
  Cross-correlating $XX$ and $XY$  pairs is equivalent to replacing one of the X legs in $XX$-$XX$ cross-correlation with Y. This results in one of the terms in the bracket above to disappear and thus the result is exactly half of Eq.~\ref{eq:prim_bispec}. Cross-correlating the $XX$ and $YY$  estimator, we get the same terms as we do as in Eq.~\ref{eq:prim_bispec}. However, in this case, the two integrals are not equal and thus we have to keep them both
\barr
  \langle \hat{\kappa}_{XX}(\bll) \hat{\kappa}_{YY}(\bll') \rangle_B &=& \Bigg[ \intLp F_{YY}(\bl_3, \bl_4) \mathcal{B}_{\bll, \bl_3, \bl_4}^{\kappa g^Y g^Y} + \intL F_{XX}(\bl_1, \bl_2) \mathcal{B}_{\bll', \bl_1, \bl_2}^{\kappa g^X g^X} \Bigg] (2\pi)^2 \delta(\bll + \bll') \, ,
  \earr
  where $g^{X}$ and $g^{Y}$ represent the foreground interlopers for line $X$ and $Y$ respectively. $XY$-$XY$ estimator cross-correlation does not produce this term.
  \item \textbf{Secondary bispectrum bias to $C_L^{\hat{\kappa} \hat{\kappa}}$:} this term has 4 components, one of which has been shown in Tab.~\ref{tab:all_terms}. This is equivalent to applying the lensing quadratic estimator with one leg coming from $t(\bl_1)$ and the other one from $t(\bl_3)$. The lensing weights $F_{XX}(\bl_1, \bl_2)$ and $F_{XX}(\bl_3, \bl_4)$ however have been determined to optimize the quadratic estimator for $\bl_1, \bl_2$ and $\bl_3, \bl_4$ pairs respectively. Thus the $\kappa$ reconstructed this way with inefficient weights is suboptimal. Similar to Case B, this $\kappa$ is correlated with the low redshift interloper line and this gives rise to the bispectrum. As the lensing reconstruction is not the optimal one, this bispectrum is called the secondary bispectrum given as
  \beq
  \bal
  \langle \hat{\kappa}_{XX}(\bll) \hat{\kappa}_{XX}(\bll') \rangle_C &= \intL \intLp F_{XX}(\bl_1, \bl_2) F_{XX}(\bl_3, \bl_4) \big\langle \big(t(\bl_1) t(\bl_3) \big) \big(g(\bl_2) g(\bl_4)\big) \big\rangle_c (2\pi)^2 \delta(\bll + \bll') \\
  &= \intL \intLp F_{XX}(\bl_1, \bl_2) F_{XX}(\bl_3, -\bll - \bl_3) 
  \Big[f_{XX}(\bl_1, \bl_3) \mathcal{B}_{\bl_1 + \bl_3, \bl_2, -\bll - \bl_3}^{\kappa g g} \\
  &\hspace{7cm}+ f_{XX}(\bl_1, \bl_4) \mathcal{B}_{\bl_1 + \bl_4, \bl_2, -\bll - \bl_4}^{\kappa g g}\\
  &\hspace{7cm}+ f_{XX}(\bl_2, \bl_3) \mathcal{B}_{\bl_2 + \bl_3, \bl_2, -\bll - \bl_3}^{\kappa g g}\\
  &\hspace{7cm}+ f_{XX}(\bl_2, \bl_4) \mathcal{B}_{\bl_2 + \bl_4, \bl_1, -\bll - \bl_4}^{\kappa g g} \Big] \, ,
  \eal
  \eeq
  where we make use of Eq.~\ref{eq:limavg}. This term does not arise for $XX$-$YY$ cross-correlation. For $XX$-$XY$ case, we have 
  \barr
  \langle \hat{\kappa}_{XX}(\bll) \hat{\kappa}_{XY}(\bll') \rangle_C
  = \intL \intLp F_{XX}(\bl_1, \bl_2) F_{XY}(\bl_3, -\bll - \bl_3) && \Big[ f_{XY}(\bl_1, -\bll - \bl_3) \mathcal{B}_{\bl_1 -\bll - \bl_3, \bl_2, \bl_3}^{\kappa g^{X} g^{X}} \nonumber \\
  && + f_{XY}(\bl_2, -\bll - \bl_3) \mathcal{B}_{\bl_2 - \bll - \bl_3, \bl_1, \bl_3}^{\kappa g^{X} g^{X}} \Big] \, ,
  \earr
  and the $XY$-$XY$ case, this simply changes to
   \barr
  \langle \hat{\kappa}_{XY}(\bll) \hat{\kappa}_{XY}(\bll') \rangle_C
  = \intL \intLp F_{XY}(\bl_1, \bl_2) F_{XY}(\bl_3, -\bll - \bl_3) && \Big[ f_{XX}(\bl_1, \bl_3) \mathcal{B}_{\bl_1 + \bl_3, \bl_2, -\bll - \bl_3}^{\kappa g^{Y} g^{Y}} \nonumber \\
  && + f_{YY}(\bl_2, -\bll - \bl_3) \mathcal{B}_{\bl_2 - \bll - \bl_3, \bl_1, \bl_3}^{\kappa g^{X} g^{X}} \Big] \, .
  \earr
  Fig.~\ref{fig:biases} shows that the secondary bispectrum term is much larger for $\hat{\kappa}_{XX}\hat{\kappa}_{XY}$ than for $\hat{\kappa}_{XX}\hat{\kappa}_{XX}$ and $\hat{\kappa}_{XY}\hat{\kappa}_{XY}$.
 This can be understood as follows.
Intuitively, one would expect the primary and secondary bispectrum terms to have similar order of magnitude.
However, the secondary bispectrum turns out much smaller than the primary for $\hat{\kappa}_{XX}\hat{\kappa}_{XX}$, because the four terms inside the secondary bispectrum turn out to cancel two-by-two almost exactly.
When considering instead $\hat{\kappa}_{XX}\hat{\kappa}_{XY}$ or $\hat{\kappa}_{XY}\hat{\kappa}_{XY}$, only two of the four terms are present, which breaks the cancellation only for $\hat{\kappa}_{XX}\hat{\kappa}_{XY}$.
  \item \textbf{Trispectrum bias to $C_L^{\hat{\kappa} \hat{\kappa}}$:} since the density field is non-Gaussian, the foreground interlopers have a non-zero trispectrum which gives rise to a bias to $C_L^{\hat{\kappa} \hat{\kappa}}$. This term's contribution to Eq.~\ref{eq:bispec_ensemble} becomes
  \beq
  \langle \hat{\kappa}_{XX}(\bll) \hat{\kappa}_{XX}(\bll') \rangle_D = \intL \intLp F_{XX}(\bl_1, \bl_2) F_{XX}(\bl_3, \bl_4) \mathcal{T}_{\bl_1, \bl_2, \bl_3, \bl_4}^{g g g g}(2\pi)^2  \delta(\bll+\bll') \, ,
  \eeq
  where $\big\langle \big(g(\bl_1) g(\bl_2)\big) \big(g(\bl_3) g(\bl_4)\big) \big\rangle_c = \mathcal{T}_{\bl_1, \bl_2, \bl_3, \bl_4}^{g g g g}$ is the trispectrum of the interloper line. We do not have this term for the $XX$-$YY$, $XX$-$XY$, and $XY$-$XY$ estimator cross-correlations as the foregrounds for line $X$ and $Y$ do not lie at the same redshift. 
\end{itemize}

\section{Signal-to-noise ratio for the lensing power spectrum}
\label{app:snr_lensing_cross}

We have shown that the cross-correlation $C_l^{\hat{\kappa}_{\rm XY} \hat{\kappa}_{\rm CMB}}$, is immune to interloper biases. 
Here, we describe the procedure and assumptions we follow to calculate the signal-to-noise ratio (SNR) on $C_l^{\hat{\kappa}_{\rm XY} \hat{\kappa}_{\rm CMB}}$.
If the fields $\hat{\kappa}_{\rm XY}$ and $\hat{\kappa}_{\rm CMB}$ were Gaussian, the standard formula for the SNR of a cross-spectrum would apply:
\beq
\Bigg( \frac{S}{N} \Bigg)^2 
= 
\sum_{{l_b}_{\rm min}}^{{l_b}_{\rm max}} (2{l_b}+1) \: f_{\rm sky} \: \Delta l  
\frac{(C_{l_b}^{\hat{\kappa}_{\rm XY} \hat{\kappa}_{\rm CMB}})^2}
{(C_{l_b}^{\hat{\kappa}_{\rm XY} \hat{\kappa}_{\rm CMB}})^2 
+ 
C_{l_b}^{\hat{\kappa}_{\rm XY}} C_{l_b}^{\hat{\kappa}_{\rm CMB}}} \, ,
\label{eq:gaussian_snr_cross_spectrum}
\eeq
with 
\beq
C_{l_b} = \frac{1}{\Delta l} \sum_{l \in [l_1, l_2]} C_l \, ,
\eeq
where $\Delta l$ is the bin width and $f_{\rm sky}$ is the sky fraction observed. 
The auto-spectra $C_{l_b}^{\hat{\kappa}_{\rm XY}}$ and $C_{l_b}^{\hat{\kappa}_{\rm CMB}}$ include the lensing signals, their noise biases and any potential additional biases. 
In other words, 
\beq
\bal
&C_{l}^{\hat{\kappa}_{\rm XY}} = C_{l}^{\kappa_{\rm XY}} + N^0_\text{LIM} + \mathcal{B}^p + \mathcal{B}^s + \mathcal{T} \\
&C_{l}^{\hat{\kappa}_{\rm CMB}} = C_{l}^{\kappa_{\rm CMB}} + N^0_\text{CMB}\ ,
\eal
\label{eq:lensing_auto_spectra}
\eeq
where $N^0$ is the Gaussian reconstruction noise given by Eq.~\ref{eq:variance}, and $\mathcal{B}^p$, $\mathcal{B}^s$, and $\mathcal{T}$ are the primary and secondary bispectrum biases and trispectrum biases respectively as shown in App.~\ref{app:biases}. 
In practice, CMB lensing from Simons Observatory and CMB-S4 will be signal-dominated on the scales we consider here ($L \lesssim 1000$), such that the reconstruction noise $N^0_\text{CMB}$ is negligible.
Here are the assumptions we make while calculating the SNR this way:
\begin{itemize}
  \item We follow Eq.~\eqref{eq:gaussian_snr_cross_spectrum} to compute the SNR, implicitly assuming that the reconstructed lensing fields $\hat{\kappa}_{\rm XY}$ and $\hat{\kappa}_\text{CMB}$ are Gaussian.
  In reality, this is not the case, since they are quadratic in the data (CMB or LIM), and in the case of the LIM, the data itself is non-Gaussian.
  This should lead to mode coupling between the various $L$-bins, whereas Eq.~\eqref{eq:gaussian_snr_cross_spectrum} only includes the diagonal elements of the covariance matrix. 
  However, we do include some of these terms, as we explain now.
  \item The interloper foregrounds present in LIM $X$ and $Y$ do not bias the cross-spectrum with CMB lensing; however, they do bias the auto-spectrum of $\hat{\kappa}_\text{XY}$, which contributes to the covariance matrix in Eq.~\eqref{eq:gaussian_snr_cross_spectrum} via the terms $\mathcal{B}^p$, $\mathcal{B}^s$, and $\mathcal{T}$ in Eq.~\eqref{eq:lensing_auto_spectra}.
  In other words, the LIM-pair estimator successfully nulls the interloper lensing bias, but not the interloper lensing noise.
  We do include these terms in the calculation.
  \item As described in \cite{Foreman18, Schaan18}, the fact that the target lines $X$ and $Y$ themselves are non-Gaussian causes additional bias and noise, similar to the term $\mathcal{B}^p$, $\mathcal{B}^s$, and $\mathcal{T}$.
  We neglect these terms here and throughout the paper, assuming that the bias-hardening method of \cite{Foreman18} allows to reduce them.
  \item Like the target lines, the interloper foregrounds are lensed as well, giving rise to a lensed foreground term \cite{Mishra19, Schaan18}. 
  This term does not bias the measured cross spectrum, but acts as an additional source of noise. We neglect this term here.
  \item Finally, like in CMB lensing, higher order noise biases $N^{(i)}$ contribute to the noise on the cross-spectrum, potentially lowering its SNR \cite{Bohm20}. Evaluating these terms is beyond the scope of this paper, and we therefore neglect them.
\end{itemize}
Therefore, with all these assumptions, the SNR we calculate can be considered an upper limit within the configuration we have considered.

\end{document}